\documentclass[a4paper, 11pt]{article}

\RequirePackage[OT1]{fontenc}
\usepackage{amsfonts,amssymb,graphics,epsfig,verbatim,bm,latexsym,amsmath,url,amsbsy,multirow,rotating,tikz,enumerate}
\usepackage[textheight=722pt,textwidth=500pt,centering,headheight=50pt,headsep=12pt,footskip=12pt]{geometry}
\usepackage[colorlinks,citecolor=blue,linkcolor=purple!60!black,urlcolor=purple!60!black]{hyperref}
\usepackage{amsmath, amstext, amsfonts, xcolor}
\usepackage{pgfrcs}
\usepackage{pgfkeys}
\usepackage{pgfmath}
\usepackage{graphicx,psfrag,epsf}
\usepackage{float}
\usepackage{subfig}
\usepackage{url}
\usepackage{lscape}
\usepackage{multirow,rotating}
\usepackage[framemethod=TikZ]{mdframed}
\usepackage[multiple]{footmisc}
\usepackage{subfloat}
\usepackage{comment}
\usepackage{booktabs}
\usepackage{enumitem}
\usepackage{subfig}
\usepackage{url}
\usepackage{mathtools}
\usepackage{caption}
\usepackage{rotating}
\usepackage{lscape}
\usepackage{float}
\usepackage{makecell}
\usepackage{tablefootnote}

\allowdisplaybreaks

\begin{document}

\begin{center}
\LARGE \bfseries{Educational Note: Paradoxical Collider Effect in the Analysis of Non-Communicable Disease Epidemiological Data: a reproducible illustration and web application}
\end{center}

{\Large
\begin{center}
Miguel Angel Luque-Fernandez*
\footnote{Biomedical Research Institute. Non-Communicable and Cancer Epidemiology Group (ibs.Granada), Andalusian School of Public Health, University of Granada, Spain}\textsuperscript{,}\footnote{Department of Non-Communicable Disease Epidemiology. London School of Hygiene and Tropical Medicine, U.K.}\textsuperscript{,}\footnote{Centre de Recherche en Epidemiologie, Biostatistique et Recherche Clinique Ecole de Sante Publique, Universite Libre de Bruxelles, Belgium}\textsuperscript{,}\footnote{Department of Epidemiology. Harvard School of Public Health. Harvard University. Boston, MA, USA}\textsuperscript{,}\footnote{Biomedical Network Research Centers of Epidemiology and Public Health (CIBERESP), ISCIII, Madrid, Spain}

Michael Schomaker
\footnote{Centre of Infectious Disease Epidemiology and Research. University of Cape Town, Cape Town, South Africa}

Daniel Redondo-Sanchez \textsuperscript{1,}\hspace{-5pt} \textsuperscript{5}

Maria Jose Sanchez Perez \textsuperscript{1,}\hspace{-5pt} \textsuperscript{5}

Anand Vaidya
\footnote{Brigham and Women's Hospital. Harvard Medical School, Harvard University, Boston, MA, USA.}\\

Mireille E. Schnitzer
\footnote{Faculty of Pharmacy and Department of Social and Preventive Medicine, University of Montreal, Montreal, Canada}\textsuperscript{,}\footnote{Department of Epidemiology, Biostatistics and Occupational Health. McGill University, Montreal, Canada}

\end{center}
}

\hspace{-20pt} * \textbf{Corresponding author}\\
Miguel Angel Luque-Fernandez\\
Biomedical Research Institute of Granada (ibs.Granada),\\
Non-communicable disease and Cancer Epidemiology Group,\\
University of Granada, Spain.\\
Campus Universitario de Cartuja,\\
C/Cuesta del Observatorio 4, 18080 Granada, Spain.\\
Tel.: +34 958 027400.\\
\href{mailto:miguel.luque.easp@juntadeandalucia.es}{miguel.luque.easp@juntadeandalucia.es}

\section*{Abstract}
Classical epidemiology has focused on the control of confounding but it is only recently that epidemiologists have started to focus on the bias produced by colliders. A collider for a certain pair of variables (e.g., an outcome Y and an exposure A) is a third variable (C) that is caused by both. In a directed acyclic graph (DAG), a collider is the variable in the middle of an inverted fork (i.e., the variable C in A $\rightarrow$ C $\leftarrow$ Y). Controlling for, or conditioning an analysis on a collider (i.e., through stratification or regression) can introduce a spurious association between its causes. This potentially explains many paradoxical findings in the medical literature, where established risk factors for a particular outcome appear protective. We use an example from non-communicable disease epidemiology to contextualize and explain the effect of conditioning on a collider. We generate a dataset with 1,000 observations and run Monte-Carlo simulations to estimate the effect of 24-hour dietary sodium intake on systolic blood pressure, controlling for age, which acts as a confounder, and 24-hour urinary protein excretion, which acts as a collider. We illustrate how adding a collider to a regression model introduces bias. Thus, to prevent paradoxical associations, epidemiologists estimating causal effects should be wary of conditioning on colliders. We provide R-code in easy-to-read boxes throughout the manuscript and a GitHub repository  (\href{https://github.com/migariane/ColliderApp}{https://github.com/migariane/ColliderApp}) for the reader to reproduce our example. We also provide an educational web application allowing real-time interaction to visualize the paradoxical effect of conditioning on a collider: \href{http://watzilei.com/shiny/collider/}{http://watzilei.com/shiny/collider/}

\section*{Keywords}
epidemiological methods, causality, noncommunicable disease epidemiology.

\section*{Key messages box}
\begin{mdframed}[roundcorner=10pt, backgroundcolor=black!10, linecolor=black!5]
\begin{itemize}
\item Paradoxical associations between an outcome and exposure are common in epidemiological studies using observational data.
\item A collider is a variable that is causally influenced by two other variables.
\item Controlling for a collider in multivariable regression analyses can introduce a spurious association between its causes (e.g. exposure and outcome). 
\item Directed Acyclic Graphs based on existing subject matter knowledge can help to identify colliders. 
\item Whether or not it is advisable to adjust for a collider depends on the main analytical objective. For instance, a predictive model may condition on a collider to increase prediction accuracy, while one should typically not condition on it when estimating causal effects to prevent bias.
\end{itemize}
\end{mdframed}

\section{Introduction}
\noindent During the last 30 years, classical epidemiology has focused on the control of confounding \cite{Greenland2001}. It is only recently that epidemiologists have started to focus on the bias produced by colliders in addition to confounders \cite{Cole2009, Vander2009}. Directed acyclic graphs (DAGs) can help to visualize the assumed structural relationships between the variables under analysis. With this framework, we can distinguish between biases resulting from i) not conditioning on common causes of exposure and outcome (unadjusted confounding)  or ii) conditioning on common effects (collider bias) \cite{Hernan2004, Robins2000}. Epidemiologists use DAGs to determine the set of variables that are necessary to control for confounding and to summarize the subject-matter knowledge of the data-generating process. Using the DAGs terminology, variables, including A (exposure) and Y (outcome), are “nodes” connected by an arrow (a.k.a directed edge) and a “path” is a way to get from one node to another, traveling along its arrows. The directed arrow ($\rightarrow$) from A to Y means that one does not exclude the possibility that A causes Y \cite{Rohrer2017, Pearl1995}.\newline

\noindent A collider for a certain pair of variables (e.g. outcome and exposure) is a third variable that is caused by both of them. In DAG terminology, a collider is the variable in the middle of an inverted fork (i.e., variable C in A $\rightarrow$ C $\leftarrow$ Y) \cite{Rohrer2017, Pearl1995}. Using regression to control for a collider, or stratifying the analysis with respect to a collider, can introduce a spurious association between its causes, which can potentially introduce non-causal associations between the exposure and the outcome. This has been used to explain why the medical literature contains many paradoxical findings, where established risk factors appear protective for the outcome  \cite{Luque2016, HernandezDiaz2006, Banack2013, Whitcomb2009}. For instance, numerous studies have reported a paradoxical protective effect of maternal cigarette smoking during pregnancy on preeclampsia, which has been named the pre-eclampsia smoking paradox. This paradox is due to gestational age at delivery, which is a collider between smoking (exposure) and pre-eclampsia (outcome) \cite{Luque2016}. However, the magnitude of the resulting bias will depend on the associations between the collider and the two parent variables. \newline

\noindent We hope that this methodological note will contribute to the increasing awareness of “colliders” and an understanding of the potential magnitude of collider bias among applied epidemiologists. The remainder of this note is structured as follows: 
\begin{enumerate}[label=\roman*.]
    \item We review terminology related to DAGs and the rules one can follow to determine whether a causal effect is estimable;
    \item We demonstrate the statistical structure of collider bias using a simulated dataset;
    \item We illustrate the effect of conditioning on a collider using a realistic non-communicable disease epidemiology example (hypertension and dietary sodium intake);
    \item We provide R-code in easy-to-read boxes throughout the manuscript  and in a GitHub repository:\\ \href{https://github.com/migariane/ColliderApp}{https://github.com/migariane/ColliderApp}; and
    \item We provide readers with an educational web application allowing real-time interaction to visualize the paradoxical effect of conditioning on a collider \href{http://watzilei.com/shiny/collider/}{http://watzilei.com/shiny/collider/}.
\end{enumerate}

\section{Statistical structure of confounding and a collider effects}

\subsection{Review of confounding}

\noindent Confounding arises from common causes of the exposure (A) and the outcome (Y). Note that in Figure 1A, both the outcome (Y) and the exposure (A) share a common “parent” (direct cause). Y and A are both called “descendants” of W as they are both caused by W. The confounder wholly or partially accounts for the observed association of the exposure (A) on the outcome (Y). The presence of a confounder can lead to “confounding bias” and thus inaccurate estimates of the effect of A on Y. More precisely, bias means that the associational measure, for example, the crude odds ratio, is different from the causal effect, such as the true marginal causal odds ratio (we give a clear definition of a marginal causal effect further below). \newline

\begin{figure}[H]
\begin{center}
	\includegraphics[scale=0.31]{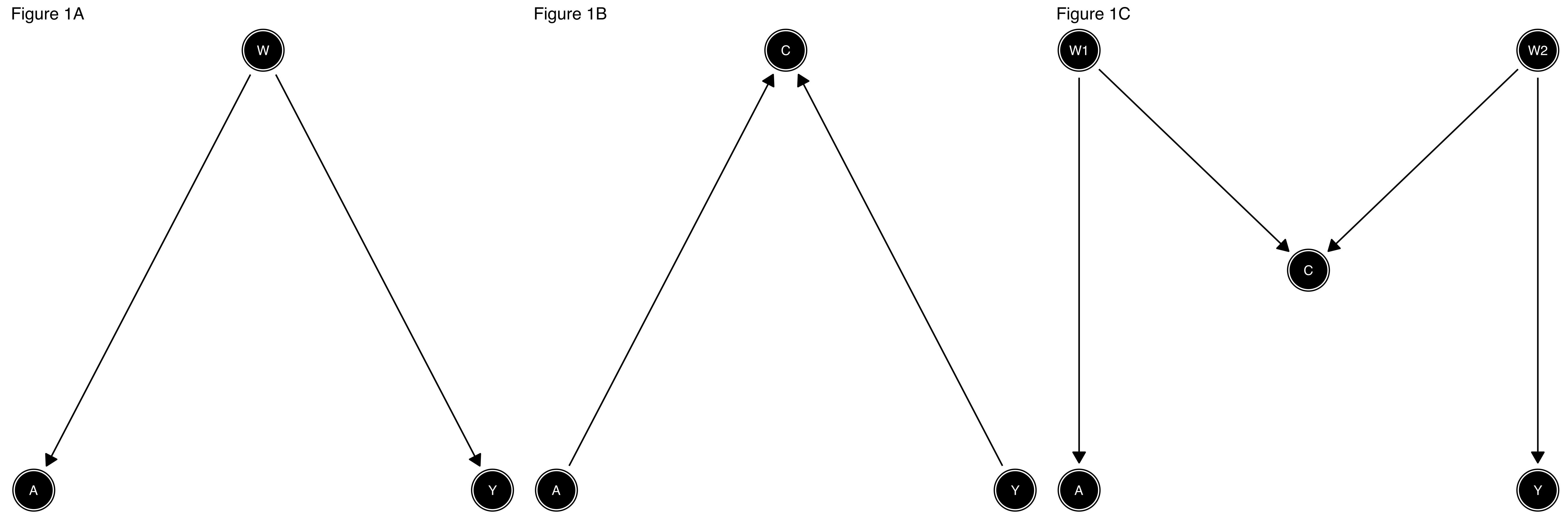}
	\caption{Basic structural associations between exposure and outcome: confounding (A), collider (B), and M-bias (C).}
	\label{figure:Figure1}
\end{center}
\end{figure}

\noindent Figure 1A gives an example of a confounding structure, where the path A $\rightarrow$ W $\leftarrow$ Y is called a “back-door path” which is defined as any path from A to Y that starts with an arrow into A. Without conditioning on variables, a path is open when it does not contain colliders. An open back-door path can be blocked, and confounding removed, by conditioning on non-colliders (via regression or stratification). In Figure 1A, conditioning on the confounder W blocks the open back-door path.  A path that is blocked by a collider can be opened by conditioning on the collider  \cite{Whitcomb2009}. To sufficiently control for confounding, epidemiologists must identify a set of variables in the DAG that block all open back-door paths from the exposure (A) to the outcome (Y) by conditioning on variables along each path (i.e., using stratification or regression). In statistical terms, being able to block all back-door paths is known as conditional exchangeability or ignorability.\newline

\noindent To describe confounding and collider bias we may use the expression association is not causation. This means that measures of association, such as the conditional mean difference in the case of a binary A, E(Y$\mid$A=1,W) - E(Y$\mid$A=0,W), is not identical to its marginal causal counterpart, the average treatment effect: E(Y(1))-E(Y(0)). Causal effects are often formulated in terms of potential outcomes, as formalised by Rubin \cite{Rubin2005}. Let A denote a continuous exposure, W a pre-exposure vector of potential confounders, and Y a continuous outcome. Each individual has a potential outcome corresponding to any given level of the exposure, that is, the outcome they would have received had they been exposed to A=a, denoted Y(a). However, it is only possible to observe a single realisation of the outcome for an individual. We may observe Y(a) only for those who were exposed with A=a \cite{Rubin2005}. If W is the set of confounding variables, then Y(a)$\perp$A$\mid$W refers to conditional exchangeability, where the symbol $\perp$ means “independent”. It implies that (within the strata of W) the distribution of Y(a) is the same regardless of the value of A that the individual actually received, i.e. E(Y(a)$\mid$A, W). We, therefore, have no systematic differences in how subjects would have performed under any given exposure that are not already explained by W.

\noindent

\subsection{Demonstration of confounding and regression adjustment}

\noindent We now demonstrate adjustment for confounding via linear regression models. In Box 1 we show how to generate data consistent with the DAG from Figure 1A after which we run two different regression models. The confounder W is generated as a standard normal random variable i.e. with mean 0 ($\mu=0$) and variance 1 ($\sigma^2=1$). The generation of A depends on the value of W plus an error term and Y is generated depending on both A and W plus an error term, where both error terms have independent standard normal distributions. Note that the simulation assumes linear relationships between the variables and that the true simulated causal effect of the exposure A on Y is 0.3 (the coefficient in the linear regression model). Then, we fit unadjusted (fit1) and adjusted (fit2: adjusted for W) linear regression models to estimate associations between A and Y. We visualize the fit of both models using the R software package visreg, where we used R version 3.5.1 (R Foundation for Statistical Computing, Vienna, Austria).\newline

\textbf{Box 1}
\begin{mdframed}[roundcorner=10pt, backgroundcolor=black!10, linecolor=black!5]
\small\begin{verbatim}
library(visreg) # load package to visualize regression output
library(ggplot2)# load package to visualize regression output
N <- 1000       # sample size 
set.seed(777)
W <- rnorm(N)                           # confounder
A <- 0.5 * W + rnorm(N)                 # exposure
Y <- 0.3 * A + 0.4 * W + rnorm(N)       # outcome
fit1 <- lm(Y ~ A)                       # crude model  
fit2 <- lm(Y ~ A + W)                   # adjusted model
# visualize crude and adjusted models
visreg(fit1, "A", gg = TRUE, line = list(col = "blue"),
points = list(size = 2, pch = 1, col = "black")) + theme_classic()
visreg(fit2, "A", gg = TRUE, line = list(col = "blue"),
points = list(size = 2, pch = 1, col = "black")) + theme_classic()
\end{verbatim}
\end{mdframed}

\noindent Note that our confounder W is the only variable that does not have parents in Figure 1A, i.e., it is not caused by any variable in the DAG. Therefore, in the code, it is the only variable that is generated independently of the other variables in the model. However, both A and Y depend on a common cause W (their parent) which is the source of the open back-door path between A and Y. As an illustration of the confounding bias due to W, Table 1 (columns 1, 2) shows the coefficients of A and W from the fitted regression models. The first regression does not condition on W and therefore has an upwards bias in the coefficient of A (0.471). However, the second regression closes the open back-door path by including the confounder W in the regression model. Thus, it estimates the causal effect as 0.289, close to the true coefficient (0.3) (Figure 2A, Table 1: columns 1, 2), the residual difference being entirely due to sampling variability.

\begin{figure}[ht!]
	\begin{center}
		\includegraphics[scale=0.47]{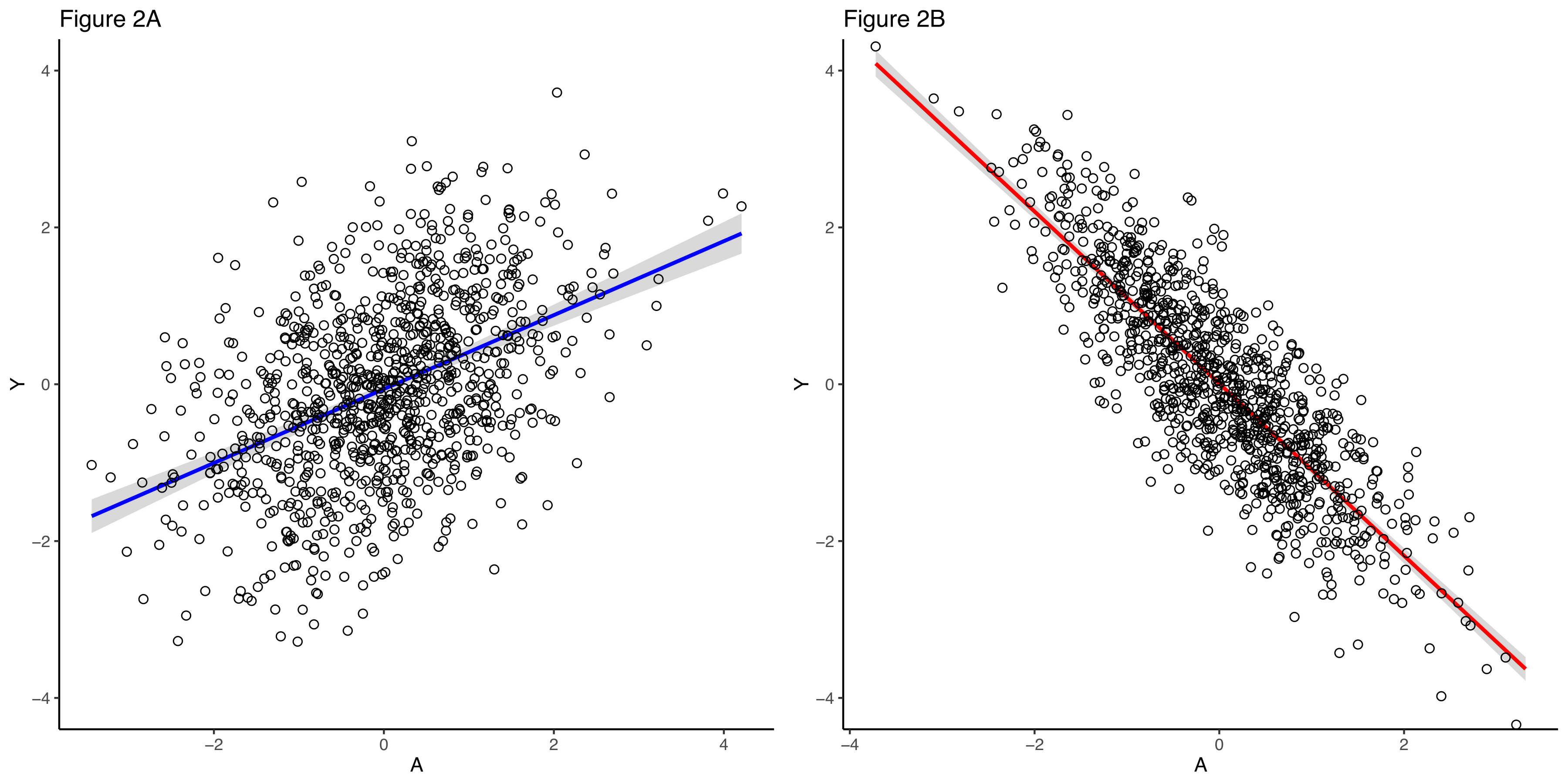}
		\caption{Visualization of the collider effect: Figure 2A model fit2 (Box 1) and Figure 2B model fit4 (Box 2).}
		\label{figure:Figure2}
	\end{center}
\end{figure}

\subsection{Collider structure}

Unlike in Figure 1A, where the causal arrows start from W, in Figure 1B they now point towards C from A and Y. If we condition on C (e.g. using regression or stratification), we will create collider bias. The common effect C is referred to as a collider on the path A $\rightarrow$ C $\leftarrow$ Y because two arrow heads collide on this node. For intuition, suppose that rain (A) and a sprinkler (Y) are the only two causes of a wet ground (C). We also assume that the sprinkler is on a daily timer, and not related to the weather. Then, if the ground is wet, knowing that it hasn’t rained implies that the sprinkler must be on. If we ignore the colliding structure, we may conclude that rain has a negative effect on the sprinkler even when we know a priori that this is not the case \cite{Pearl2009}. \newline

\noindent Conditioning on the collider induces an association between the potential outcomes (Y(a)) and the exposure (A) and conditional ignorability (Y(a) $\perp$ A $\mid$ W, C) no longer holds. In other words: in Figures 1B and 1C, conditioning on the collider C opens the back-door path between A and Y which was previously blocked by the collider itself (A $\leftarrow$ C $\rightarrow$ Y). Thus, the association between A and Y would be a mixture of the association due to the effect of A on Y and the association due to the open back-door path. Thus, association would not be causation anymore.

\noindent Figure 1C gives another, more complex, collider structure, usually known as M-bias, in which the collider (C) is the effect of a common cause (W1) of the exposure (A) and a common cause (W2) of the outcome (Y). There is only one back-door path, and it is already blocked by the collider (C) thus we do not need to control for anything. This is the difference between confounders and colliders: a path will be open if one does not adjust for confounders, but blocked if adjustment is made; for colliders, it is the other way around.  However, some could consider C to be a classical confounder as it is associated with both A, via (A $\leftarrow$ W1 $\rightarrow$ C), and with Y, via a path that does not go through A (C $\leftarrow$ W2 $\rightarrow$ Y), and it is not in the causal pathway between A and Y. However, controlling for C will introduce a collider bias. Note that if you use the traditional characteristics used to identify confounders (i.e., a third variable (W) associated with both the exposure (A) and the outcome (Y) that is not in the causal pathway between A and Y) one can confuse a collider with a confounder.

\noindent To simulate the scenario portrayed in Figure 1B, we generate data, again using a simple linear data generating mechanism (Box 2). First, we simulate A as a standard normally distributed variable. Y equals the value of A plus an error term and C is generated depending on both A and Y, plus error. Note that as shown in Figure 1B, now the exposure A and the outcome Y, are the parents of C (their common effect). We fit the unadjusted model excluding the collider (fit3) and then the model including the collider (fit4: collider model). The true causal coefficient of the exposure A is -1.2, and the coefficients for the association of the collider C with the exposure A and the outcome Y are 1.0 and 1.0, respectively (Box 2). \newline

\newpage

\textbf{Box 2}
\begin{mdframed}[roundcorner=10pt, backgroundcolor=black!10, linecolor=black!5]
\small\begin{verbatim}
library(visreg)  # load package to visualize regression output 
library(ggplot2) # load package to visualize regression output
N <- 1000        # sample size
set.seed(777)
A <- rnorm(N)                       # exposure
Y <- -1.2 * A + rnorm(N)             # outcome
C <- 1 * A + 1 * Y + rnorm(N)       # collider
fit3 <- lm(Y ~ A)                   # crude model
fit4 <- lm(Y ~ A + C)               # adjusted model
# visualize crude and adjusted models
visreg(fit3, "A", gg = TRUE, line = list(col = "red"),
points = list(size = 2, pch = 1, col = "black")) + theme_classic()
visreg(fit4, "A", gg = TRUE, line = list(col = "red"),
points = list(size = 2, pch = 1, col = "black")) + theme_classic()
\end{verbatim}
\end{mdframed}

\noindent Table 1 (columns 3, 4) shows the coefficient of A in the unadjusted model (fit3) and the coefficients of A and C in the model adjusting for the collider (fit4). Unlike in the previous section, the simpler regression without C approximately recovers the true coefficient of A (0.3) with an estimate of 0.326, while the regression adjusting for C is substantially biased (-0.416). The model which includes the collider (fit4) is not unequivocally inferior from a predictive point of view, where the main focus is to improve the model’s predictive performance. For instance, the model containing the collider has a much lower Akaike Information Criterion (AIC) than the one without the collider (Table 1). However, conditioning on the collider C has paradoxically changed the direction of the association between A and Y (Figure 2B, Table 1: column 3). Thus, in this case, conditioning on the collider in the regression model introduces a bias while ignoring the collider does not add bias. The paradoxical negative association occurs when both A and Y are positively correlated with the collider.\newline

\noindent From this demonstration, it is clear that subject-matter knowledge (i.e., plausible biological mechanisms in clinical epidemiological settings) is necessary to perform causal estimation  \cite{Hernan2002}. Thus, using DAGs to communicate causal structural relationships between variables helps in identifying variables that act as a collider, and identify where conditioning may create non-causal associations between the exposure (A) and outcome (Y) \cite{Hernan2002, Greenland1999, Pearce2014}.

\section{Motivating Example}\label{sec:motivation}

\subsection{Data generation}

\noindent Based on a motivating example in non-communicable disease epidemiology, we generated a dataset with 1,000 observations to contextualize the effect of conditioning on a collider. Nearly 1 in 3 Americans suffer from hypertension and more than half do not have it under control \cite{Benjamin2017}. Increased levels of systolic blood pressure over time are associated with increased cardiovascular morbidity and mortality \cite{Gu2008}.\newline

\noindent Summative evidence shows that exceeding the recommendations for 24-hour dietary sodium intake in grams (gr) is associated with increased levels of systolic blood pressure (SBP) in mmHg \cite{Sacks2001}. Furthermore, with advancing age, the kidney undergoes several anatomical and physiological changes that limit the adaptive mechanism responsible for maintaining the composition and volume of the extracellular fluid. These include a decline in glomerular filtration rate and the impaired ability to maintain water and sodium homeostasis in response to dietary and environmental changes \cite{Tareen2005}. Likewise, age is associated with structural changes in the arteries and thus SBP \cite{Gu2008}. \newline

\noindent Age is a common cause of both high SBP and impaired sodium homeostasis. Thus, age acts as a confounder for the association between sodium intake and SBP (i.e. age is on the back-door path between sodium intake and SBP) as depicted in Figure 3. However, high levels of 24-hour excretion of urinary protein (proteinuria) are caused by sustained high SBP and increased 24-hour dietary sodium intake. Therefore, as depicted in Figure 3, proteinuria acts as a collider  (via the path SOD $\rightarrow$ PRO $\leftarrow$ SBP). In a realistic scenario one might control for proteinuria if physiological factors influencing SBP are not completely understood by the researcher, the relationships between variables are not depicted in a DAG or proteinuria is conceptualized as a confounder. Controlling for proteinuria (PRO) introduces collider bias.  \newline

\begin{figure}[ht!]
	\begin{center}
		\includegraphics[scale=0.7]{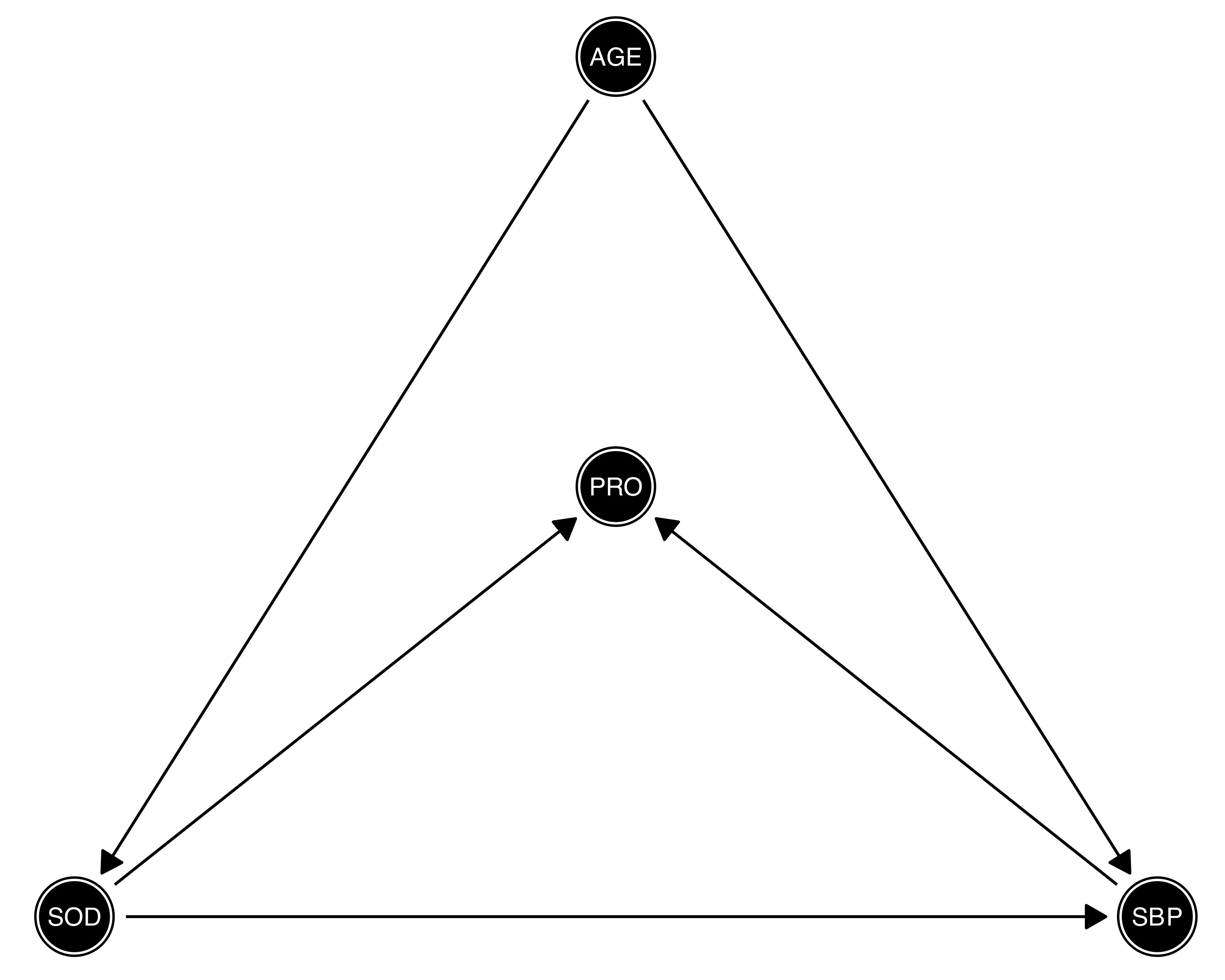}
		\caption{Directed acyclic graph depicting the structural causal relationship of the exposure and outcome, confounding and collider effects. Exposure: 24-hour sodium dietary intake in gr (SOD), outcome: systolic blood pressure in mmHg (SBP), confounder: age in years (AGE), collider: 24-hour urinary protein excretion, proteinuria (PRO).}
		\label{figure:Figure3}
	\end{center}
\end{figure}

\noindent We are interested in estimating the effect of 24-hour dietary sodium intake (in grams) on SBP, adjusting for age. The objective of the illustration is to show the paradoxical effect of 24-hour dietary sodium intake on SBP after conditioning on a collider (proteinuria). Box 3 shows the data generation for the simulated data based on the structural relationship between the variables depicted in the DAG from Figure 3. We assumed that SBP is a common cause of age and dietary sodium intake. We also simulated 24-hour excretion of urinary protein as a function of age, SBP, and sodium intake. We aimed to have a range of values of the simulated data which was biologically plausible and as close to reality as possible  \cite{Van2016, Carroll2000}.  \newline

\noindent Supplementary Table 1 shows the descriptive statistics (minimum, maximum, mean, median, first and third quartiles) of the generated data. Note that for educational purposes, we present the code and results for a single dataset simulated by our data-generating mechanism. However, at the end of the illustration, we also present the results of 1,000 Monte-Carlo simulations with a sample size of 10,000 patients aiming to quantify the bias associated with conditioning on a collider.  \newline

\noindent The simulation assumes linear relationships between the variables. Thus, the interpretation of the beta coefficients in the formulae of the code on Box 3 is straightforward. The true causal effect of sodium intake on SBP is 1.05 (i.e., $ \text{Systolic blood pressure}\, = \,\beta_{1}\,\times\,\text{sodium}\,+\,\beta_{2}\,\times\,\text{age}\,+\,\varepsilon$; where $\beta_1= 1.05$, $\beta_2 = 2.00$ and $\varepsilon$ is a standard normally distributed error). The coefficients for the association of PRO with SBP and sodium intake are 2.0 and 2.8, respectively (i.e., $ \text{Proteinuria}\, = \,\beta_{1}\,\times\,\text{SBP}\,+\,\beta_{2}\,\times\,\text{Sodium}\,+\,\varepsilon$; where $\beta_1= 2.0$, $\beta_2 = 2.8$ and $\varepsilon$ is a standard normally distributed error) (Box 3). Supplementary Figure 1 shows the functional form for each variable and the multivariable Spearman’s correlation matrix.  \newline

\begin{figure}[H]
	\begin{center}
		\includegraphics[scale=0.47]{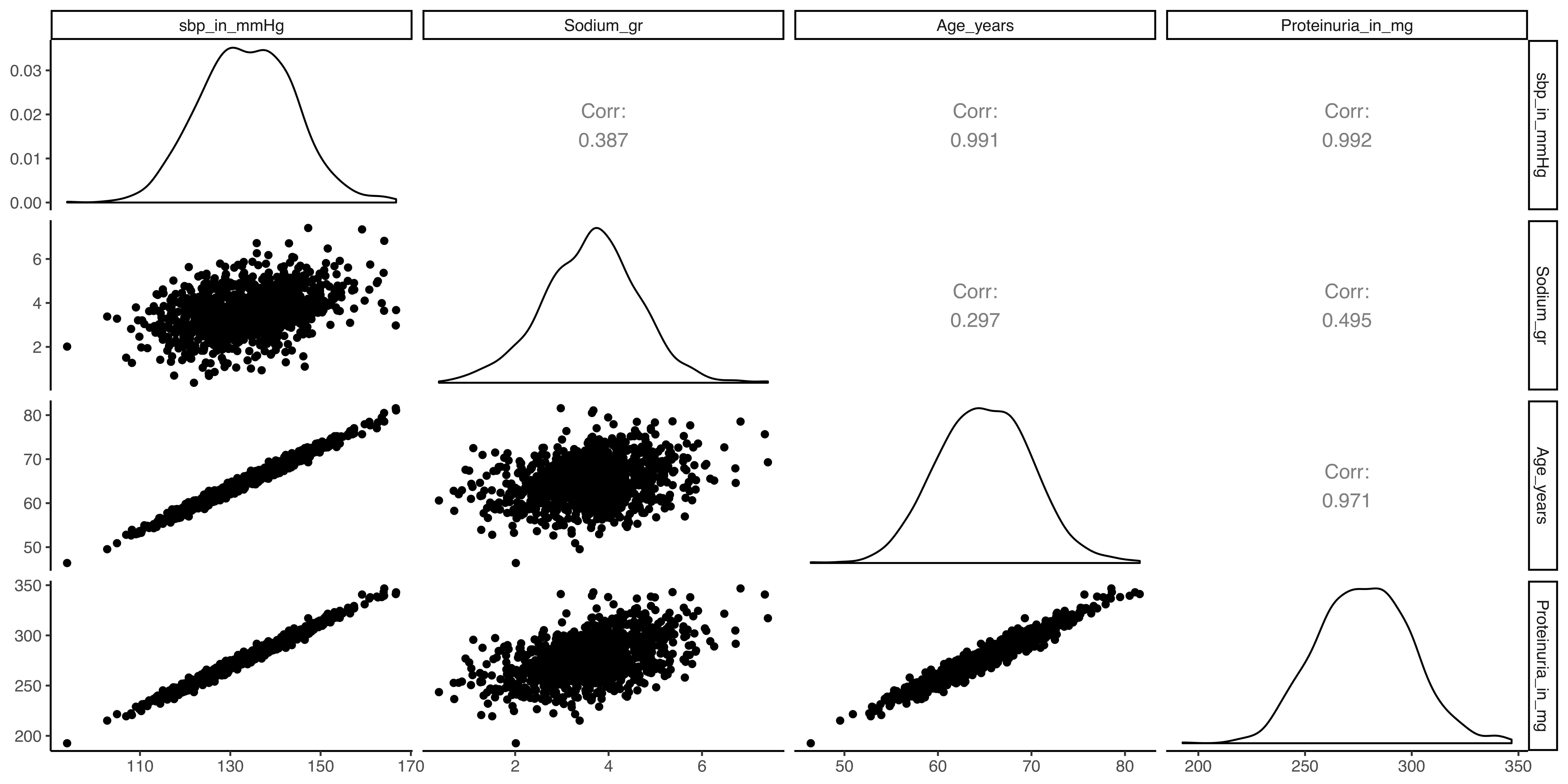}
		\caption*{\textbf{Supplementary Figure 1}: Visualization of the multivariate structure of the data generation, n = 1,000.}
		\label{figure:Supplementary Figure 1}
	\end{center}
\end{figure}

\textbf{Box 3}
\begin{mdframed}[roundcorner=10pt, backgroundcolor=black!10, linecolor=black!5]
\small\begin{verbatim}
generateData <- function(n, seed){
set.seed(seed)
Age_years <- rnorm(n, 65, 5)
Sodium_gr <- Age_years / 18 + rnorm(n) 
sbp_in_mmHg <- 1.05 * Sodium_gr + 2.00 * Age_years + rnorm(n)
hypertension <- ifelse(sbp_in_mmHg>140,1,0)
Proteinuria_in_mg <- 2.00*sbp_in_mmHg + 2.80*Sodium_gr + rnorm(n)
data.frame(sbp_in_mmHg, hypertension, Sodium_gr, Age_years, Proteinuria_in_mg) 
}
ObsData <- generateData(n = 1000, seed = 777)
\end{verbatim}
\end{mdframed}

\noindent We fit three different linear regression models (Box 4) to evaluate the effect of sodium intake on SBP: i) unadjusted model, ii) model adjusted for age, iii) model adjusted for age and the collider (proteinuria). The model specifications are shown here below; in Box 4 we show how to fit and visualize the corresponding models in R. \newline

\textit{Models specification:}

$$\text{Systolic Blood Pressure in mmHg} = \beta_{0}\,+\,\beta_{1}\,\times\,\text{Sodium in gr} \,+\, \varepsilon$$
$$\text{Systolic Blood Pressure in mmHg} = \beta_{0}\,+\,\beta_{1}\,\times\,\text{Sodium in gr}\,+\,\beta_{2}\,\times\,\text{Age in years}\,+\, \varepsilon $$
$$ \text{Systolic Blood Pressure in mmHg} = \beta_{0}\,+\,\beta_{1}\,\times\,\text{Sodium in gr}\,+\,\beta_{2}\,\times\,\text{Age in years}\,+\,\beta_{3}\times\,\text{Proteinuria in mg} \,+\, \varepsilon $$

\newpage

\textbf{Box 4}
\begin{mdframed}[roundcorner=10pt, backgroundcolor=black!10, linecolor=black!5]
\small\begin{verbatim}
library(broom) # load packages to visualize regression model’s output
library(visreg)
## Models Fit
fit0 <- lm(sbp_in_mmHg ~ Sodium_gr, data = ObsData);tidy(fit0)
fit1 <- lm(sbp_in_mmHg ~ Sodium_gr + Age_years , data = ObsData);tidy(fit1)
fit2 <- lm(sbp_in_mmHg ~ Sodium_gr + Age_years + Proteinuria_in_mg, data = ObsData);
tidy(fit2)

## Models visualization
par(mfrow = c(1,3))
visreg(fit0, ylab = "SBP in mmHg", line = list(col = "blue"),
points = list(cex = 1.5, pch = 1), jitter = 10, bty = "n")
visreg(fit1, ylab = "SBP in mmHg", line = list(col = "blue"),
points = list(cex = 1.5, pch = 1), jitter = 10, bty = "n")
visreg(fit2, ylab = "SBP in mmHg", line = list(col = "red"),
points = list(cex = 1.5, pch = 1), jitter = 10, bty = "n")
\end{verbatim}
\end{mdframed}

\noindent We also fit three logistic regression models to evaluate the effect of sodium intake on hypertension defined as a binary outcome (SBP $\geq$ 140 mmHg = 1, SBP $<$140 mmHg = 0): i) an unadjusted model, ii) a model adjusted for age, and iii) a model adjusted for age and the collider (proteinuria). The model specifications are the same as described above but now with a binary outcome (hypertension); in Box 5 we show how to fit and visualize the corresponding models in R using a forest plot function.\newline

\textbf{Box 5}
\begin{mdframed}[roundcorner=10pt, backgroundcolor=black!10, linecolor=black!5]
\small\begin{verbatim}
## Models fit on multiplicative scale
library(dplyr) 
library(forestplot)
fit3 <- glm(hypertension ~ Sodium_gr, family=binomial(link='logit'), data=ObsData)
or <- round(exp(fit3$coef)[2],3) # conditional odds ratio from logistic model
ci95 <- exp(confint(fit3))[-1,]  # 95% CI of odds ratio

fit4 <- glm(hypertension ~ Sodium_gr + Age_years, family = binomial(link = "logit"), 
data = ObsData) 
or <- round(exp(fit4$coef)[2],3)
ci95 <- exp(confint(fit4))[2,]

fit5 <- glm(hypertension ~ Sodium_gr + Age_years + Proteinuria_in_mg, 
family = binomial(link = "logit"),
        data = ObsData) 
or <- round(exp(fit5$coef)[2],3)
ci95 <- exp(confint(fit5))[2,]

## Forest plot (see supplementary material for accessing the complete code)
fp <- rbind(result1,result2,result3);fp %>% or_graph()
\end{verbatim}
\end{mdframed}

\subsection{Effect of conditioning on a collider}\label{sec:collider}

\noindent Table 2 shows the model coefficients and goodness of fit from the linear regression models. Figure 4 shows the regression line and 95\% confidence interval for the predicted level of SBP, illustrating the effect of conditioning on a collider. The adjusted regression line was derived as the predicted estimate of SBP, conditional on the median value of age for Figure 4B and age and proteinuria for Figure 4C \cite{Breheny2017}. As opposed to the unadjusted and bivariate models (Figures 4A and 4B), the collider model (Figure 4C) suggests a negative relationship between sodium intake and SBP (i.e., for one unit increase in sodium intake, the expected SBP decreases by 0.9 mmHg). Figure 5 shows odds ratios from the logistic regression models. The odds ratio for the effect of sodium on hypertension similarly suggests that it is protective (i.e., for one unit increase in sodium intake the risk of hypertension decreases by 98\%) (Figure 5). 

\begin{figure}[H]
	\begin{center}
		\includegraphics[scale=0.47]{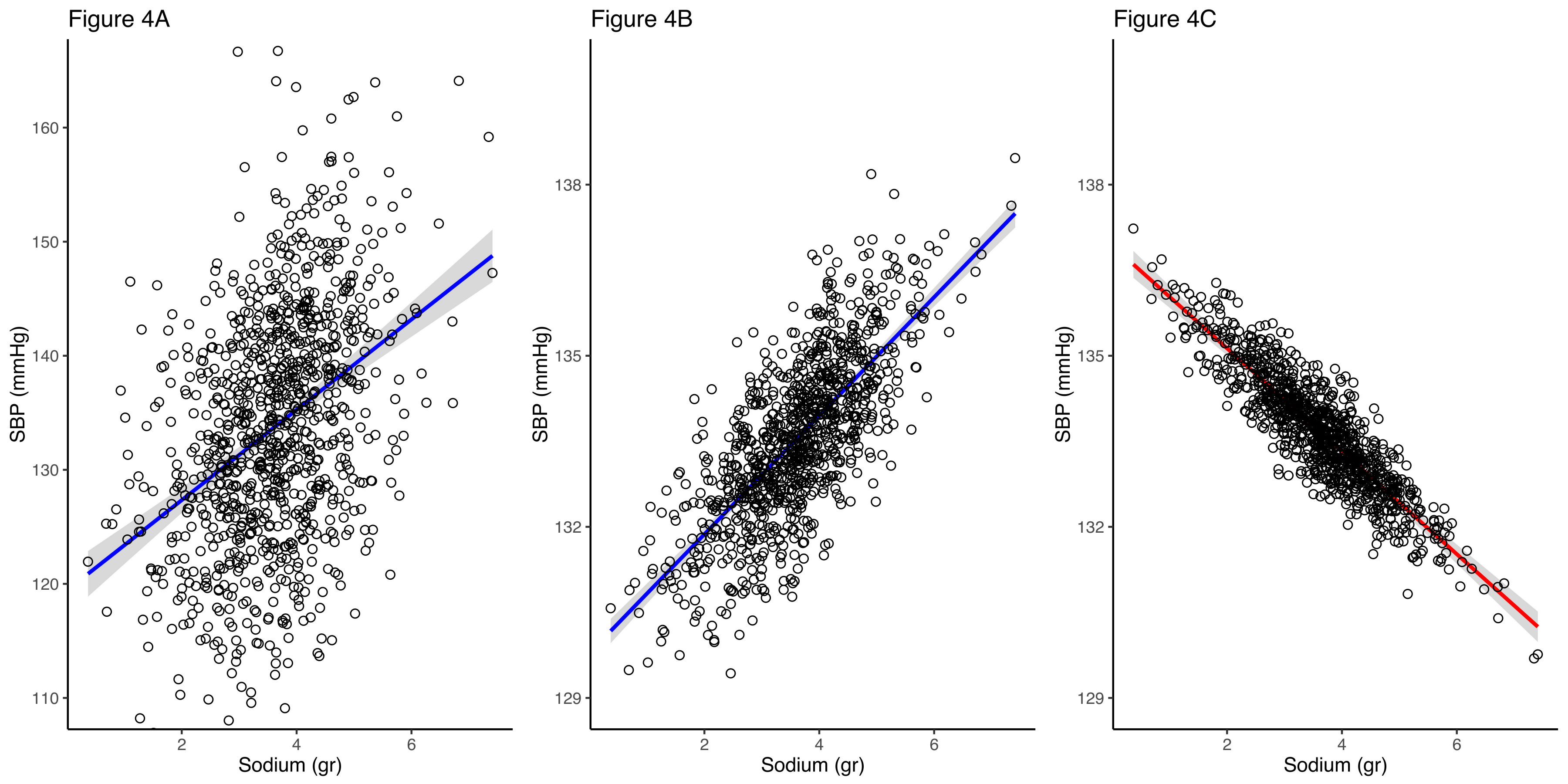}
		\caption{Collider effect for the illustration: Univariate (Figure 4A), bivariate (Figure 4B) and multivariate (Figure 4C) coefficients and standard errors for the linear association between systolic blood pressure and 24-hour sodium dietary intake adjusted for age acting as a confounder and proteinuria acting as a collider, n = 1,000.}
		\label{figure:Figure 4}
	\end{center}
\end{figure}

\begin{figure}[H]
	\begin{center}
		\includegraphics[scale=0.42]{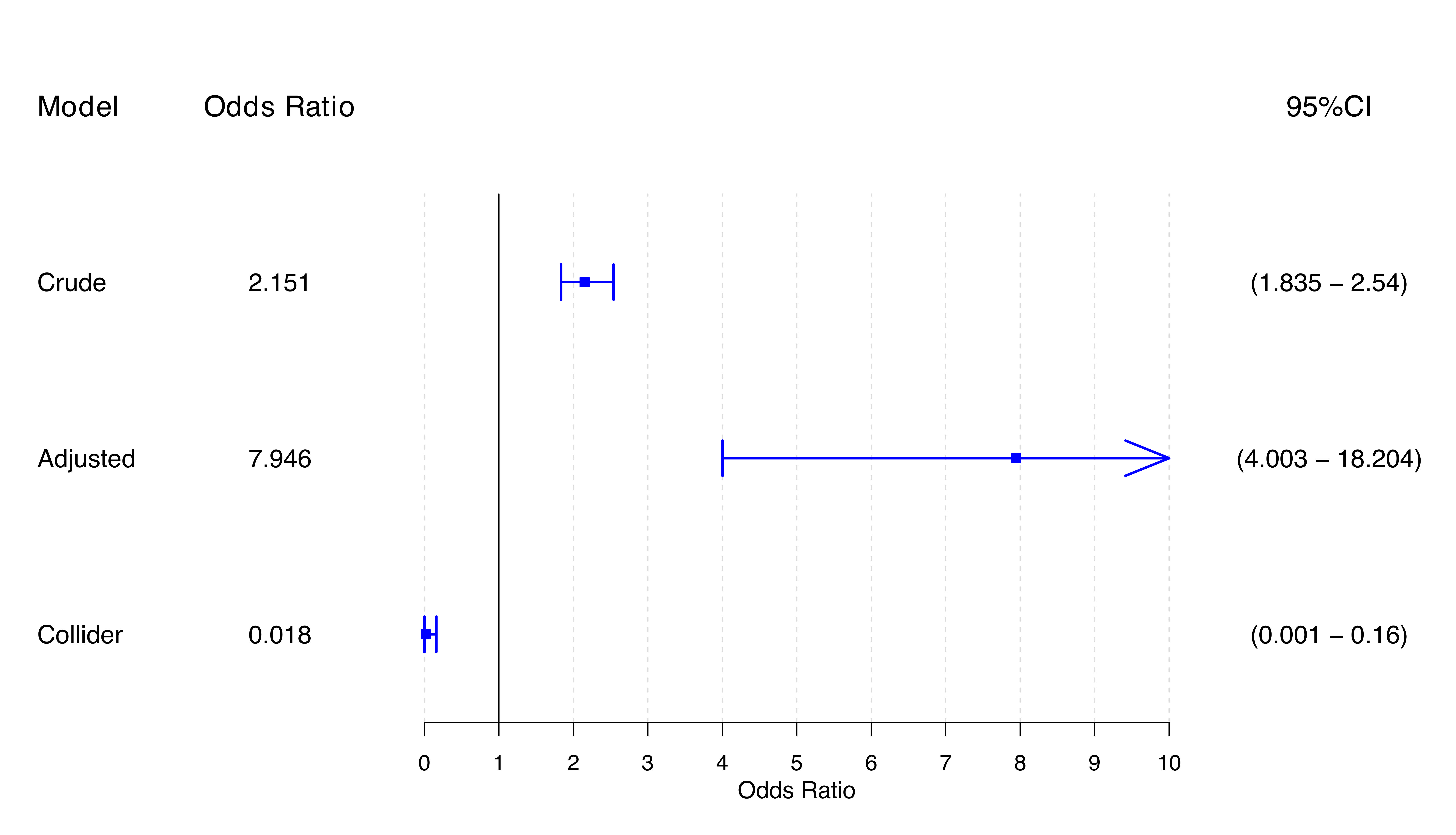}
		\caption{Collider effect for the illustration in a multiplicative scale for the effect of 24-hour sodium dietary intake on systolic blood pressure adjusted for age acting as a confounder and proteinuria acting as a collider, n = 1,000.}
		\label{figure:Figure 5}
	\end{center}
\end{figure}

\subsection{Monte-Carlo Simulation Results}

\noindent Box 6 shows the code used to run the Monte-Carlo simulation on the additive scale using the same setting as in Box 3. The true simulated causal effect of 24-hour sodium intake on SBP was 1.05 mmHg in the linear model and the coefficients for the association of PRO with SBP and sodium intake were 2.0 and 2.8, respectively. After 1,000 simulation runs,  the estimated additive effect of 24-hour sodium intake on SBP was -0.91 mmHg (i.e., for one unit increase in sodium intake there was a decrease of -0.91 units in SBP). The relative bias due to conditioning on proteinuria (the collider) was 13.3\%.   \newline

\textbf{Box 6}
\begin{mdframed}[roundcorner=10pt, backgroundcolor=black!10, linecolor=black!5]
\small\begin{verbatim}
# Monte Carlo Simulations
R<-1000
true <- rep(NA, R)
collider <- rep(NA,R)
se <- rep(NA,R)
set.seed(050472)
for(r in 1:R) {
if (r%%10 == 0) cat(paste("This is simulation run number", r, "\n"))

# Function to generate data 
generateData <- function(n){
Age_years <- rnorm(n, 65, 5)
Sodium_gr <- Age_years / 18 + rnorm(n)
sbp_in_mmHg <- 1.05 * Sodium_gr + 2.00 * Age_years + rnorm(n)
Proteinuria_in_mg <- 2.00 * sbp_in_mmHg + 2.80 * Sodium_gr + rnorm(n)
data.frame(sbp_in_mmHg, Sodium_gr, Age_years, Proteinuria_in_mg)
}
ObsData <- generateData(n=10000) 

# True effect
true[r] <- summary(lm(sbp_in_mmHg ~ Sodium_gr + Age_years, data = ObsData))$coef[2,1]

# Collider effect
collider[r] <- summary(lm(sbp_in_mmHg ~ Sodium_gr + Age_years + Proteinuria_in_mg,
               data = ObsData))$coef[2,1]
se[r] <- summary(lm(sbp_in_mmHg ~ Sodium_gr + Age_years + Proteinuria_in_mg,
               data = ObsData))$coef[2,2]
}

# Estimate of sodium true effect 
mean(true)
# Estimate of sodium biased effect in the model including the collider
mean(collider)
# simulated standard error/confidence interval of outcome regression
lci <- (mean(collider) - 1.96*mean(se)); mean(lci)
uci <- (mean(collider) + 1.96*mean(se)); mean(uci)
# Bias 
Bias <- (true - abs(collider));mean(Bias)
# % Bias
relBias <- ((true - abs(collider)) / true); mean(relBias) * 100
# Plot bias
plot(relBias)

\end{verbatim}
\end{mdframed}

\noindent The code included in all of the boxes is provided in a supplementary file. We also provide the link to a web application \href{http://watzilei.com/shiny/collider/}{http://watzilei.com/shiny/collider/} (Supplementary Figure 2) where users can dynamically modify the values of the true causal effect and the coefficients in the data generation process of the collider model. The collider web application allows users to interactively modify the range of values of the slider input and visualize the collider effect of the example. As shown in the web application the strength of the association of the collider with both the exposure and the outcome determines the strength of the paradoxical protective effect of 24-hour diary sodium intake in gr on systolic blood pressure. \newline

\begin{figure}[H]
	\begin{center}
		\includegraphics[scale=0.34]{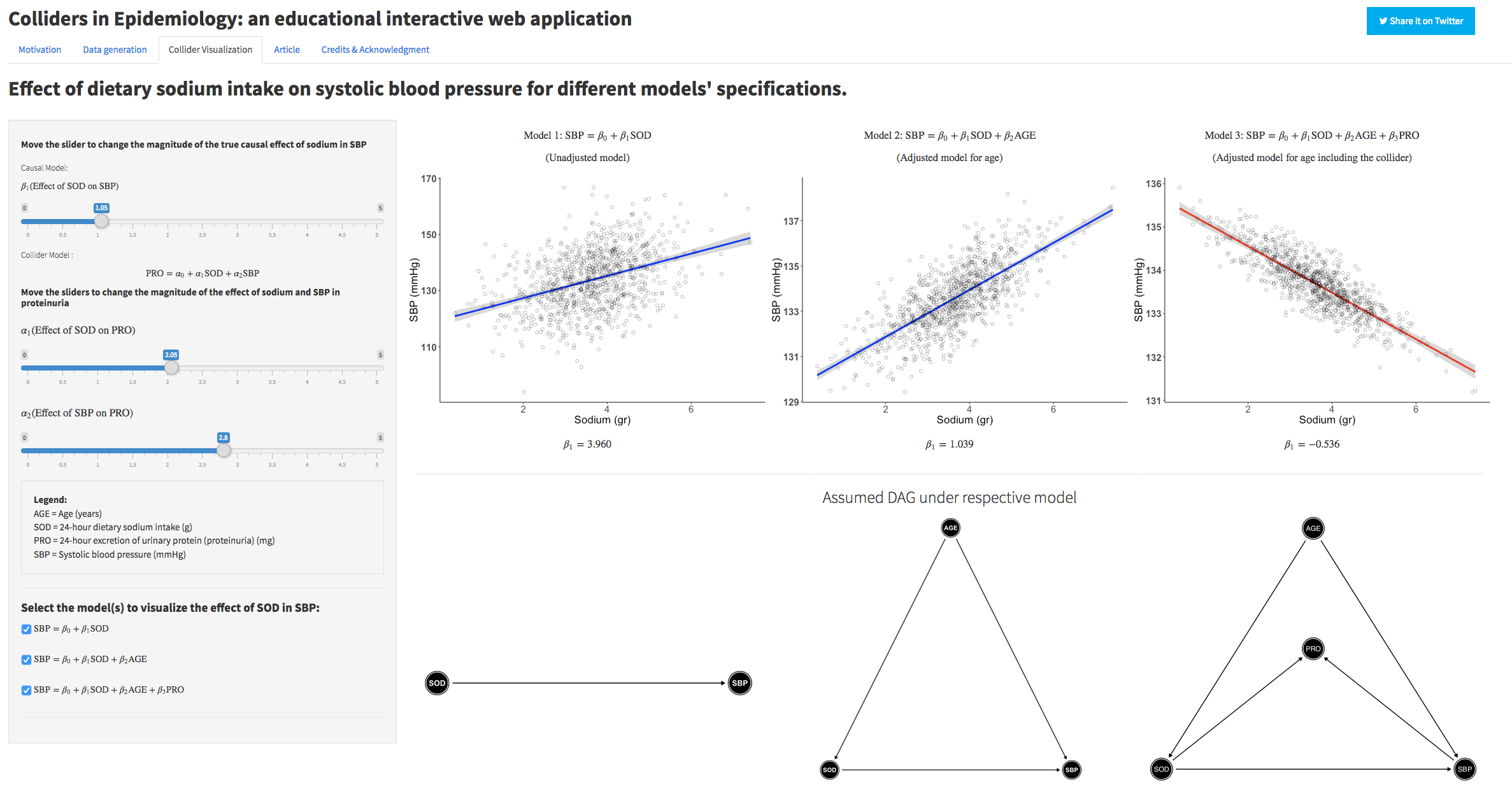}
		\caption*{\textbf{Supplementary Figure 2}: Screenshot collider Shiny web application.}
		\label{figure:Supplementary Figure 2}
	\end{center}
\end{figure}

\noindent The magnitude of the causal effect between the exposure and the outcome and the collider with the exposure and the outcome determines whether paradoxical effects arise when conditioning on the collider. Table 3 shows different values for the true causal effect of sodium intake on SBP and the estimated causal effect for different values of the association between PRO (i.e., the collider) with sodium intake ($\alpha_1$) and SBP ($\alpha_2$) in the collider model, and assuming $\alpha_1 = \alpha_2$ (i.e., the same magnitude for the collider-exposure and the collider-outcome associations in the collider model). Overall, with this data-generating structure, the collider bias reduces the magnitude of the estimated causal effect between sodium intake and SBP. To create a paradoxical effect (i.e., the negative association between sodium intake and SBP) we found that increasing the true causal effect requires an increase of the strength of the association between collider-exposure and collider-outcome association with respect to the magnitude of the true causal effect (Table 3). Note that assuming $\alpha_1 = \alpha_2$ is not realistic, but it is a convenient simplification that helps to gain intuition about changes in the magnitude of bias. \newline

\noindent There are two additional situations where collider bias arises that are important to point out: i) collider bias arising not from the choice of variables to control in the analysis, but from conditioning on a measured or unmeasured common effect of the exposure and the outcome in a sample selection; ii) situations where the collider is both a collider and a confounder \cite{Greenland2003}. \newline

\noindent Recent evidence shows that even modest influences on sample selection can generate biased and potentially misleading estimates of both phenotypic and genotypic associations \cite{Munafo2017}. However, the solution is not often clear as information regarding sample selection and attrition might be unmeasured. On the other hand, in M-bias settings where the collider is also a confounder, it is useful to understand the tradeoffs in bias between collider and confounder control. The size in the magnitude of collider bias may often be comparable in size with bias from classical confounding \cite{Greenland2003}. It has been shown that M-bias has a small impact unless associations between the collider and confounders were very large (relative risk$>$8). Generally, in this situation, controlling for confounding would be prioritized over avoiding M-bias \cite{Liu2012}. \newline
\section{Conclusion}\label{sec:conclusion}

We investigated a situation where adding a certain type of variable to a linear regression model, called a “collider”, led to bias with respect to the regression coefficient estimates while still improving the model fit. DAGs are based on subject matter knowledge and are vital for identifying colliders. Determining if a variable is a collider involves critical thinking about the true unobserved data generation process and the relationship between the variables for a given scenario \cite{Pearce2014,Pearce2016}. Then, the decision whether to include or exclude the variable in a regression model using observational data in epidemiology is based on whether the purpose of the study is prediction or explanation/causation. Under the structures we investigated here, adding a collider to a regression model is not advised when one is interested in the estimation of causal effects, as this may open a back-door path. However, if prediction is the purpose of the model, the inclusion of colliders in the models may be advisable if it reduces the model’s prediction error. Most research in epidemiology tries to explain how the world works (i.e., it is causal), thus to prevent paradoxical associations, epidemiologists estimating causal effects should be aware of such variables. 

\section*{Competing Interests}
The authors declare that they do not have any conflict of interest associated with this research and the content is solely the responsibility of the authors.

\section*{Funding}
Miguel Angel Luque Fernandez is supported by the Spanish National Institute of Health, Carlos III Miguel Servet I Investigator Award (CP17/00206). Maria Jose Sanchez Perez is supported by the Andalusian Department of Health. Research, Development and Innovation Office project grant PI-0152/2017. Anand Vaidya was supported by the National Institutes of Health (grants DK107407 and DK115392) and by the Doris Duke Charitable Foundation (award 2015085). Mireille E. Schnitzer is supported by a New Investigator Salary Award from the Canadian Institutes of Health Research.

\section*{Authors’ contributions}
The article and Shiny application arise from the motivation to disseminate the principles of modern epidemiology among clinicians and applied researchers. MALF developed the concept, designed the study, carried out the simulation, analysed the data, and wrote the article. DRS and MALF developed the shiny application. All authors interpreted the data, and drafted and revised the manuscript, code for the manuscript, and code for the Shiny application. All authors read and approved the final version of the manuscript. MALF is the guarantor of the article.

\section{Tables}

\newpage

\begin{landscape}
\centering
\textbf{Table 1}. Coefficients and standard errors of the linear association between Y (outcome) and A (exposure) illustrating confounding and collider effects, n = 1,000

\begin{table}[!ht]
\centering

\begin{tabular}{lrrrrrr}
 & \multicolumn{1}{l}{} & \multicolumn{1}{l}{} & \multicolumn{1}{l}{} & \multicolumn{1}{l}{} & \multicolumn{1}{l}{} & \multicolumn{1}{l}{} \\ \cline{2-7}
 & \multicolumn{6}{c}{Dependent variable (Y)} \\ \cline{2-7}
 & \multicolumn{2}{c}{W (confounder)} & \multicolumn{1}{l}{} & \multicolumn{1}{l}{} & \multicolumn{2}{c}{C (collider)} \\ \cline{2-3} \cline{6-7}
 & \multicolumn{1}{l}{\makecell{Unadjusted\\ coefficient}} & \multicolumn{1}{l}{\makecell{Adjusted\\ coefficient}} & \multicolumn{1}{l}{} & \multicolumn{1}{l}{} & \multicolumn{1}{l}{\makecell{Unadjusted\\ coefficient}} & \multicolumn{1}{l}{\makecell{Adjusted\\ coefficient}} \\ \cline{2-7}
 & \multicolumn{1}{l}{(Standard Error)} & \multicolumn{1}{l}{(Standard Error)} & \multicolumn{1}{l}{} & \multicolumn{1}{l}{} & \multicolumn{1}{l}{(Standard Error)} & \multicolumn{1}{l}{(Standard Error)} \\ \hline
 & (Fit 1) & (Fit 2) &  &  & (Fit 3) & (Fit 4) \\
\textbf{A} & \textbf{0.471} & \textbf{0.289} & \textbf{} & \textbf{A} & \textbf{0.326} & \textbf{-0.416} \\
 & (-0.030) & (-0.032) &  &  & (-0.031) & (-0.035) \\
W &  & 0.425 &  & C &  & 0.491 \\
 &  & (-0.035) &  &  &  & (-0.018) \\
Intercept & -0.061 & -0.06 &  &  & 0.01 & 0.035 \\
 & (-0.033) & (-0.031) &  &  & (-0.031) & (-0.023) \\
\hline
\multicolumn{1}{l}{AIC} & 100.42 & -31.992 &  &  & -55.369 & -626.824 \\ \hline
\small{Note: Lower AIC is better} & \multicolumn{1}{l}{} & \multicolumn{1}{l}{} & \multicolumn{1}{l}{} & \multicolumn{1}{l}{} & \multicolumn{1}{l}{} & \multicolumn{1}{l}{}
\end{tabular}
\label{Table 1}
\end{table}
\end{landscape}

\begin{landscape}
\begin{table}[!ht] 
\centering
  \textbf{Table 2}.{ Univariate, bivariate and multivariate coefficients and standard errors for the linear association between systolic blood pressure and 24-hour sodium dietary intake adjusted for age acting as a confounder and proteinuria acting as a collider, n = 1,000}
\begin{tabular}{lrrrrrr}
 & \multicolumn{2}{l}{} & \multicolumn{2}{l}{} & \multicolumn{2}{l}{}  \\ \cline{2-7}
 & \multicolumn{6}{c}{Dependent variable: Systolic Blood Pressure in mmHg} \\ \cline{2-7}
 & \multicolumn{2}{l}{Univariate Coefficient} & \multicolumn{2}{l}{Bivariate Coefficient} & \multicolumn{2}{l}{Multivariate Coefficient} \\
 & \multicolumn{2}{r}{(Standard Error)} & \multicolumn{2}{r}{(Standard Error)} & \multicolumn{2}{r}{(Standard Error)} \\ \hline
 True effect of Sodium in gr: 1.05
 \\ \hline

\textbf{Sodium in gr} & & \textbf{3.960} & & \textbf{1.039} & & \textbf{-0.902} \\
& & (0.298) & & (0.032) & & (0.036) \\

Age in years & & & & 2.004  & & 0.416 \\
 & & & & (0.007)  & & (0.027) \\

Proteinuria in mg & & & &  & &  0.396  \\
 & & & & & & (0.007) \\

Intercept & & 119.420 & & -0.311 & & -0.091 \\
  & & (1.122) & & (0.407) & & (0.192) \\
\hline
AIC  & & 7363.45 & & 2807.89 & & 1302.66 \\
\hline
\small{Note: Lower AIC is better} &  & & & & & \\ 
\end{tabular}
\label{table_2}
\end{table}
\end{landscape}

\textbf{Table 3}.{ Different scenarios for the true causal effect and the magnitude of the association between the collider with the exposure ($\alpha_1$) and the outcome ($\alpha_2$), n = 1,000.}\\
{\small %
\begin{tabular}[c]{lrrr}
			\textbf{Causal model *} & \textbf{Collider model **} & & \\ \hline
			True causal effect ($\beta_1$) & \makecell{Magnitude of the association between \\ the collider with the exposure ($\alpha_1$) and \\ the outcome ($\alpha_2$), assuming $\alpha_1$ = $\alpha_2$} & Estimated causal effect & Absolute bias\\ \hline
	      1  & 0.5 & 0.630  & 0.370 \\
			 & 1.0 & 0.033 & 0.967 \\
			 & 1.5 & -0.368  &  1.368 \\
			 & 2.0 & -0.596 & 1.596 \\
			 & 2.5 & -0.727 & 1.727 \\
			 & 3.0 & -0.807 & 1.807 \\
			 & 3.5 & -0.858 & 1.858 \\
			 & 4.0 & -0.892 & 1.892 \\
			 & 4.5 & -0.916 & 1.916 \\
			 & 5.0 & -0.933 & 1.933 \\ \hline
		   2 & 0.5 & 1.453 & 0.547 \\
			 & 1.0 & 0.558 & 1.442 \\
			 & 1.5 & -0.045 & 2.045 \\
			 & 2.0 & -0.388 & 2.388 \\
			 & 2.5 & -0.586 & 2.586 \\
			 & 3.0 & -0.706 & 2.706 \\
			 & 3.5 & -0.783 & 2.783 \\
			 & 4.0 & -0.835 & 2.835 \\
			 & 4.5 & -0.871 & 2.871 \\
			 & 5.0 & -0.897 & 2.897 \\ \hline
		   3 & 0.5 & 2.277 & 0.723 \\
			 & 1.0 & 1.082 & 1.918 \\
			 & 1.5 & 0.278 & 2.722 \\
			 & 2.0 & -0.181 & 3.181 \\
			 & 2.5 & -0.445 & 3.445 \\
			 & 3.0 & -0.606 & 3.606 \\
			 & 3.5 & -0.709 & 3.709 \\
			 & 4.0 & -0.778 & 3.778 \\
			 & 4.5 & -0.826 & 3.826 \\
			 & 5.0 & -0.861 & 3.861 \\ \hline
		   4 & 0.5 & 3.100 & 0.900 \\
			 & 1.0 & 1.607 & 2.393 \\
			 & 1.5 & 0.600 & 3.400 \\
			 & 2.0 & 0.027 & 3.973 \\
			 & 2.5 & -0.304 & 4.304 \\
			 & 3.0 & -0.505 & 4.505 \\
			 & 3.5 & -0.634 & 4.634 \\
			 & 4.0 & -0.721 & 4.721 \\
			 & 4.5 & -0.781 & 4.781 \\
			 & 5.0 & -0.825 & 4.825 \\ \hline
		   5 & 0.5 & 3.923 & 1.077 \\
			 & 1.0 & 2.132 & 2.868 \\
			 & 1.5 & 0.923 & 4.077 \\
			 & 2.0 & 0.234 & 4.766 \\
			 & 2.5 & -0.163 & 5.163 \\
			 & 3.0 & -0.405 & 5.405 \\
			 & 3.5 & -0.560 & 5.560 \\
			 & 4.0 & -0.664 & 5.664 \\
			 & 4.5 & -0.737 & 5.737 \\
			 & 5.0 & -0.789 & 5.789 \\ \hline
		\label{table_3}	
\end{tabular}
\newline
\noindent \textbf{*Causal model}: SBP = $\beta_0$ + $\beta_1$ SOD + $\beta_2$ AGE + $\beta_3$PRO \\
\textbf{**Collider model}: PRO = $\alpha_0$ + $\alpha_1$SOD + $\alpha_2$SBP \\
\textbf{Absolute bias} = True – Estimate \\
AGE = Age (years)	 	 	 	  \\
SOD = 24-hour dietary sodium intake (g)	 	 	 \\
PRO = 24-hour excretion of urinary protein (proteinuria) (mg)	 	 \\
SBP = Systolic blood pressure (mmHg)
}

\newpage

\begin{landscape}
\begin{table}[!ht] \centering
\caption*{\textbf{Supplementary Table 1}: Descriptive distribution of the simulated data, n = 1,000}
\begin{tabular}{@{\extracolsep{5pt}}rrrrr}
\\[-1.8ex]\hline
\hline\\[-1.8ex]
 Systolic blood pressure in mmHg & Hypertension & Sodium in gr & Age in years & Proteinuria mg in 24h\\
\hline
Min.   : 93.86   & Min.   : 0.00 & Min.   :0.37   & Min.   :46.40   & Min.   :192.6   \\
1st Qu.:126.58   & 1st Qu.: 0.00 & 1st Qu.:2.95   & 1st Qu.:61.57   & 1st Qu.:262.2   \\
Median :133.85   & Median : 0.00 & Median :3.66   & Median :64.91   & Median :277.7   \\
Mean   :133.76   & Mean   : 0.28 & Mean   :3.62   & Mean   :65.01   & Mean   :277.7   \\
3rd Qu.:141.03   & 3rd Qu.: 1.00 & 3rd Qu.:4.26   & 3rd Qu.:68.35   & 3rd Qu.:292.8   \\
Max.   :166.73   & Max.   : 1.00 & Max.   :7.41   & Max.   :81.58   & Max.   :346.8   \\
\hline
\hline\\[-1.8ex]
Hypertension: systolic blood pressure $\geq$140 mmHg
\end{tabular}
\label{sup_table_1}
\end{table}
\end{landscape}

\bibliographystyle{vancouver}
{\footnotesize
\bibliography{collider}
}

\end{document}